\documentclass[]{emulateapj}
\def\ifm#1{\relax\ifmmode#1\else$\mathsurround=0pt #1$\fi}

\def\ltsima{$\; \buildrel < \over \sim \;$}
\def\lsim{\lower.5ex\hbox{\ltsima}}
\def\gtsima{$\; \buildrel > \over \sim \;$}
\def\gsim{\lower.5ex\hbox{\gtsima}}

\newenvironment{inlinefigure}{
\def\@captype{figure}
\noindent\begin{minipage}{0.999\linewidth}\begin{center}}
{\end{center}\end{minipage}\smallskip}

\shorttitle{Morphologies and SEDs of EROs in GOODS-South}
\shortauthors{Moustakas et al.}

\begin{document}

\submitted{Submitted to the Astrophysical Journal Letters}

\title{Morphologies and Spectral Energy Distributions of Extremely Red
Galaxies in the GOODS-South Field\altaffilmark{1}$^,$\altaffilmark{2}}

\author{Leonidas A. Moustakas\altaffilmark{3}, Stefano
  Casertano\altaffilmark{3}, Chris Conselice\altaffilmark{4}, Mark
  E. Dickinson\altaffilmark{3}, Peter Eisenhardt\altaffilmark{5},
  Henry C. Ferguson\altaffilmark{3}, Mauro Giavalisco\altaffilmark{3},
  Norman A. Grogin\altaffilmark{6}, Anton
  M. Koekemoer\altaffilmark{3}, Ray A. Lucas\altaffilmark{3}, Bahram
  Mobasher\altaffilmark{3}, Casey Papovich\altaffilmark{7},  Alvio
  Renzini\altaffilmark{8}, Rachel S. Somerville\altaffilmark{3},
  Daniel Stern\altaffilmark{5}}

\altaffiltext{1}{ 
Based on observations taken with the NASA/ESA Hubble Space Telescope,
which is operated by the Association of Universities for Research in
Astronomy, Inc.\ (AURA) under NASA contract NAS\,5--26555.}
\altaffiltext{2}{
Based on observations collected at the European Southern
Observatory, Chile (ESO Programmes 164.O-0561, 168.A-0485, 169.A-0725,
170.A-0788, 267.A-5729)} 

\altaffiltext{3}{Space Telescope Science Institute, 3700 San Martin
  Drive, Baltimore, MD  21218; {\tt leonidas, casertano, med,
    ferguson, mauro, koekemoe, somerville@stsci.edu}}
\altaffiltext{4}{California Institute of Technology, Mail Stop 105-24,
  Pasadena, CA 91109; {\tt cc@astro.caltech.edu}}
\altaffiltext{5}{Jet Propulsion Laboratory, California Institute of
  Technology, Mail Stop 169-327 (PE) \& 169-506 (DS), Pasadena, CA
  91109; {\tt prme@kromos.jpl.nasa.gov, stern@zwolfkinder.jpl.nasa.gov}}
\altaffiltext{6}{Department of Physics and Astronomy, The Johns
  Hopkins University, 3400 North Charles Street, Baltimore, MD 21218;
  {\tt nagrogin@stsci.edu}} 
\altaffiltext{7}{Steward Observatory, The University of Arizona, 933
  North Cherry Avenue, Tucson, AZ 85721; {\tt papovich@as.arizona.edu}}
\altaffiltext{8}{European Southern Observatory,
  Karl-Schwarzschild-Strasse 2, D-85748 Garching bei M\"unchen,
  Germany; {\tt renzini@eso.org}}

\begin{abstract}
Using $U'$- through $K_s$-band imaging data in the GOODS-South field,
we construct a large, complete sample of 275 ``extremely red objects''
(EROs; $K_s<22.0$, $R-K_s>3.35$; AB), all with deep \emph{HST}/ACS
imaging in $B_{435}$, $V_{606}$, $i_{775}$, and $z_{850}$, and
well-calibrated photometric redshifts.  Quantitative concentration and
asymmetry measurements fail to separate EROs into distinct
morphological classes.  We therefore visually classify the
morphologies of all EROs into four broad types: ``Early''
(elliptical-like), ``Late'' (disk galaxies), ``Irregular'' and
``Other'' (chain galaxies and low surface brightness galaxies), and
calculate their relative fractions and comoving space densities.  For
a broad range of limiting magnitudes and color thresholds, the
relative number of early-type EROs is approximately constant at
$33$--$44$\%, and the comoving space densities of Early- and Late-type
EROs are comparable.

Mean rest-frame spectral energy distributions (SEDs) at $\lambda_{\rm
rest}\approx0.1$--$1.2\mu$m are constructed for all EROs.  The SEDs
are extremely similar in their range of shapes, independent of
morphological type.  The implication is that any differences between
the broad-band SEDs of Early-type EROs and the other types are
relatively subtle, and there is no robust way of photometrically
distinguishing between different morphological types with usual
optical/near-infrared photometry.
\end{abstract}


\section{Introduction\label{sec:intro}}

The newest challenges to galaxy formation models lie in predicting or
explaining the state of galaxy assembly when the universe was in its
middle ages, at only several billion years old.  In field populations
at these redshifts, $z\sim1$--$2$, it is necessary to sample a large
enough volume at sufficient depth to overcome biases due to cosmic
variance.  This is particularly true for strongly clustered galaxies,
such as the population of ``extremely red objects'' (EROs).  These
have comoving correlation scales of $\sim\,10\,h_{100}^{-1}$\,Mpc at
$z\gsim1$ \citep{roche:02, firth:02, daddi:00a, daddi:02}, and
therefore require surveys of solid angle at least 100\,arcmin$^2$ to
have a variance less than $\sim20$\% \citep{rss:03}.

The spectral energy distributions (SEDs) of galaxies are products of
the amalgam of stellar mass functions, star formation histories, dust,
and geometry, often with degeneracies (such as in age and metallicity)
complicating interpretation.  EROs, by their simple
optical-minus-infrared color selection, embody this degeneracy
extremely well, and are known to consist of at least two distinct
classes of galaxies: old, and dusty/star-forming.  Surveys in other
wavelengths have post~facto uncovered EROs, including radio
\citep*{willott:01}, sub-mm \citep{smail:02, wehner:02}, and both
hard- and soft-X-ray surveys \citep{davo:02, brusa:02}.  Each class
(and subclass) of EROs offers a potentially important datum for
further unfolding and challenging galaxy formation models.  Currently,
predictions of CDM-based semi-analytic models do not compare well
against the numbers and statistics of EROs as a whole \citep{firth:02,
cimatti:02b}.  Having the relative fractions and space densities of,
minimally, \emph{morphologically}-determined early-type EROs versus
all EROs (e.g. \citealt*{yan:03}), should improve the situation for
comparisons against models.

Currently, there is a popular working hypothesis that some (large)
fraction of EROs \citep[those similar to the $z=1.552$
53W091,][]{spinrad:97} are the ancestors of present-day elliptical
galaxies (e.g. \citealt{daddi:00b}; \citealt*{ms:02} [MS02]; and
references therein).  Conversely, many EROs demonstrably host highly
obscured starbursts \citep[e.g. the $z=1.44$ HR10;][]{gd:96} which
could be an important component of the global luminosity density
\citep{elbaz:02, cimatti:02a}.  A photometric technique has been
proposed by \citet{pm:00} to make broad distinctions between passive
and star-forming, dusty EROs.  This is as yet not well-tested.  A
systematic morphological and photometric study of a large,
homogeneously-selected sample of EROs is therefore timely.

In this \emph{Letter} we present the selection, morphological mix, and
mean rest-frame SEDs of $K_s<22.0$ (AB) EROs, selected from the
GOODS-South \emph{HST}/ACS mosaic \citep{giav:03} over
$163$\,arcmin$^2$.  The sample selection and their redshifts are
described in \S\,\ref{sec:data}, followed by the morphological study
in \S\,\ref{sec:morph}.  In \S\,\ref{sec:results} we present the
relative fractions, space densities, and average rest-frame SEDs by
morphological type.  Conclusions are drawn in \S\,\ref{sec:conc}.  We
give magnitudes in the AB system\footnote{The most relevant
conversions to Vega magnitudes and colors for this paper are: $R_{\rm
Vega}=R-0.206$ and $K_{s,{\rm Vega}}=K_{s}-1.859$.  Thus, $R-K_s>3.35$
in AB corresponds to $(R-K_s)_{\rm Vega}>5.0$.}, and, where necessary,
we use a flat cosmology with $\Omega_m=1-\Omega_{\Lambda}=0.3$ and
$H_0=70$\,$h_{70}$\,km\,s$^{-1}$\,Mpc$^{-1}$.

\begin{inlinefigure}
\begin{center}
\resizebox{\textwidth}{!}{\includegraphics{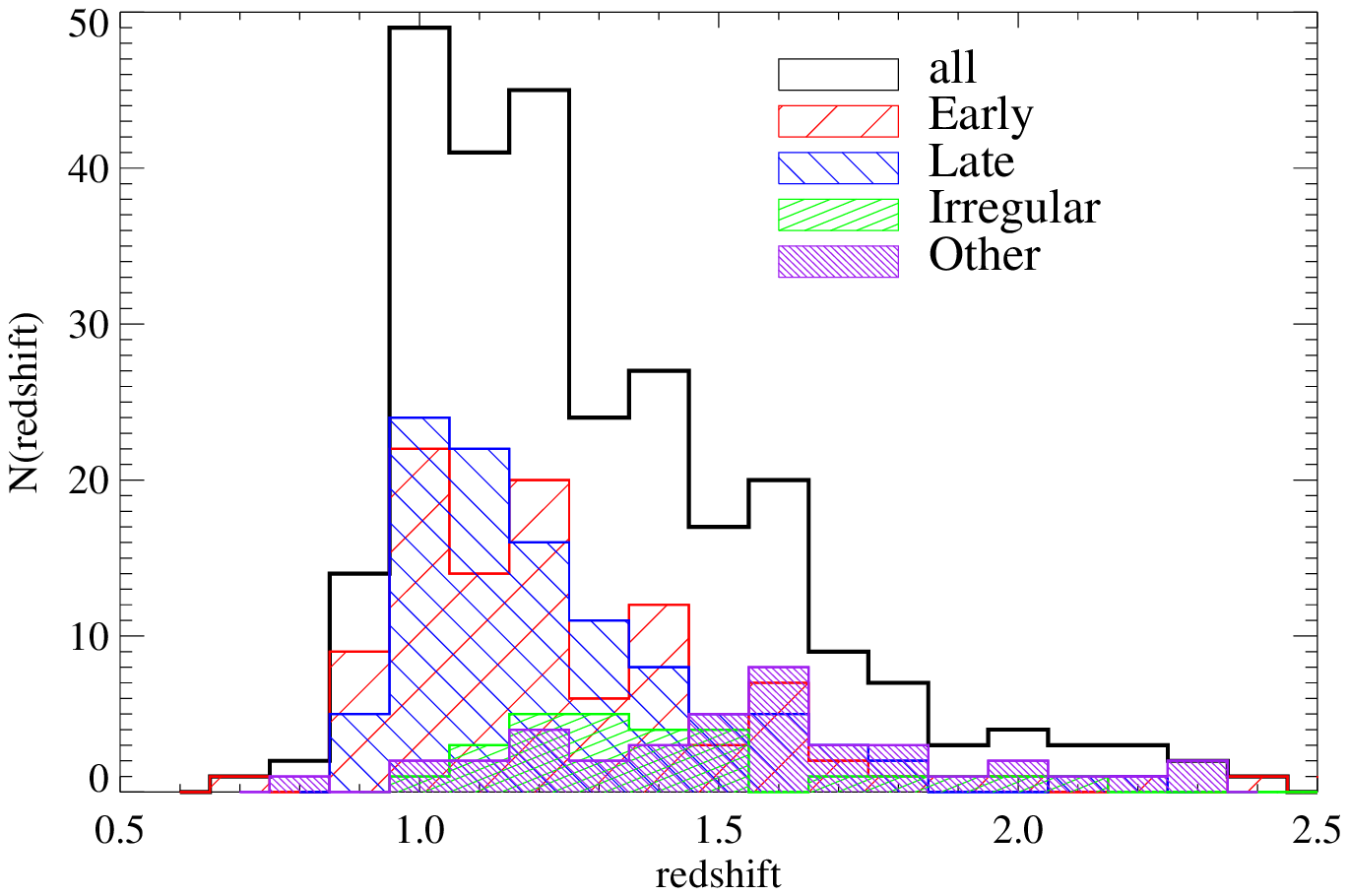}}
\figcaption{The ERO redshift distributions $N(z)$ for the full sample
($R-K_s>3.35$, $K_s<22.0$) and for each morphological type, as given
in legend. The redshift distributions for Early- and Late-types are
very similar, with $z_{\rm med}\approx1.2$, tailing to $z\sim1.5$.
The Irregular- and Other- types have a very broad $N(z)$ distribution
with $z_{\rm med}\approx1.5$, and dominate the sample at $z\gsim1.5$.
Of the 275 objects plotted here, 66 redshifts are spectroscopic and
the remainder are photometric (see text).
\label{fig:nzt}}
\end{center}
\end{inlinefigure}

\section{Sample selection and redshifts\label{sec:data}}

The GOODS data are described in \citet{giav:03}.  Briefly, for the
Southern field (overlapping with CDFS; \citealt{giac:02}), an ACS
mosaic covering $163$\,arcmin$^2$ in $B_{435}$, $V_{606}$, $i_{775}$,
and $z_{850}$ are supported by a broad set of ground-based optical and
infrared imaging, $U'UBVRIJHK_s$, from the VLT, NTT, and 2.2\,m
telescopes.  The infrared data are from the SOFI and ISAAC
instruments.  For this work, we employ a PSF-matched SOFI
$K_s$-selected catalog from which the stars have been
removed\footnote{Unresolved sources are identified by a plot of the
$z_{850}$ $0.0625$\,arcsec-diameter surface-brightness vs isophotal
magnitude, which provides a robust separation to
$z_{850}\approx26.5$.}.  The SOFI data reach $\sim50$\% completeness
at $K_{s}\approx22.8$, and are still nearly entirely complete at
$K_s\approx22.0$ \citep{moy:03, giav:03}.

We consider several definitions of EROs, all of which are essentially
complete for these survey depths: three color cuts ($R-K_s> 3.35$,
$3.65$, and $4.35$) for each of two magnitude limits ($K_s<21.0$ and
$22.0$).  The largest (``full'') sample ($R-K_s>3.35$, $K_s<22.0$)
consists of 275 objects.  The smallest sample ($R-K_s>4.35$,
$K_s<21.0$) contains 18 objects.

Photometric redshifts are computed using the method of
\citet{txitxo:00}.  These are well-calibrated to $z_{\rm
spec}\approx1.5$, with $\sigma[\Delta z/(1+z_{\rm spec})]=0.11$ (to
$R<25.5$; \citealt{mobasher:03}) using a total of 434 spectroscopic
redshifts, including 202 from a VLT/FORS2 campaign over 2002-2003 (to
be presented elsewhere), and from the K20 survey
\citep{cimatti:02b,cimatti:02c}. This sample includes 66 EROs with
$R-K_s>3.35$ and $K_s<22.0$, for which $\sigma[\Delta z/(1+z_{\rm
spec})]=0.05$.  In the analysis that follows, photometric redshifts
have been replaced with spectroscopic redshifts where possible.  There
is one faint compact ERO with a formal $z_{\rm phot}=6.4$, which is a
candidate $L$-dwarf; for completeness we keep this object in the full
sample.  The ERO redshift distribution for the full sample is shown in
Fig.~\ref{fig:nzt}.  The $N(z)$ has a fairly sharp edge at
$z\approx0.95$, a median value of $z_{\rm med}\approx1.2$.  The $N(z)$
for the redder subsets are similar, but shifted to a higher median
redshift, $z_{\rm med}\approx1.5$ (for $R-K_s>4.35$).

\begin{inlinefigure}
\begin{center}
\resizebox{\textwidth}{!}{\includegraphics{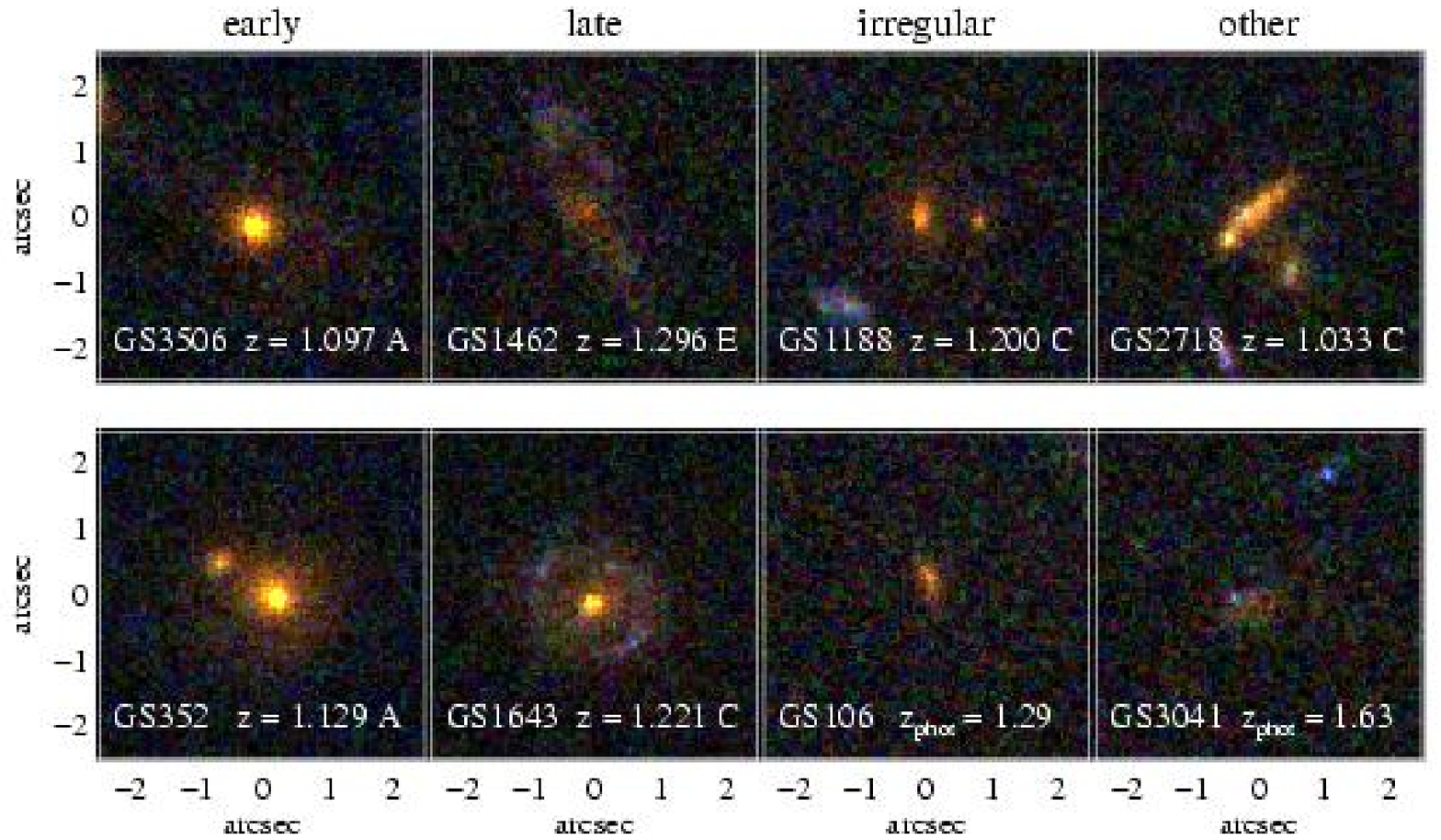}}
\figcaption{Montage of 3-color $B\,i\,z$ stamps, illustrating examples
of each morphological type, as described in the text.  Two objects are
given for each type (by column).  The caption of each stamp gives the
ID (``GS''=GOODS-South); the redshift (spectroscopic for all but for
GS\,106 and GS\,3041, for which photometric redshifts are given); and
a letter code reflecting observed spectroscopic features --
A=absorption; E=emission; C=both emission and absorption
features.\label{fig:montage}}
\end{center}
\end{inlinefigure}

\section{Morphological classification of EROs\label{sec:morph}}

The study of EROs has been largely motivated by their tantalizing
connection with both the progenitors of present-day massive elliptical
galaxies, and possibly, ultraluminous infrared galaxies.  Using the
high-resolution ACS data, it should be possible to distinguish between
these extremes morphologically.  With the ACS $z_{850}$ data,
restframe morphologies longward of 4000\AA\ can be determined for
redshifts $z\lsim1.3$, reducing possible morphological $k$-correction
issues \citep[e.g.][]{giav:96} for most EROs in this sample.  The
GOODS data reach a $z_{850}$~1-$\sigma$ isophotal limit of
$27.3$\,mag\,arcsec$^{-2}$ \citep{giav:03}.  As the average magnitude
of galaxies in the full sample is $z_{850}\approx23.0$, for most cases
the distinction between early- and non-early-types should be clear.
At redshifts beyond $z\approx1.3$, the 4000\AA-break begins to shift
past the $z_{850}$ band, and morphological $k$-corrections may become
important.  This can be solved by also examining longer-wavelength
imaging (e.g. the $\sim0.4$\,arcsec-resolution ISAAC data in the
$K_s$-band); this will be done in future work.

We began by exploring the galaxy morphologies with the concentration
and asymmetry (CAS) parameters of \citet[][see also
\citealt{abraham:96, bershady:00}]{conselice:03}.  We found that the
parameter values were fairly continuous and did not cleanly
distinguish ``classes,''
which reflects the morphological richness of all types of EROs.
We therefore
resorted to a \emph{visual} classification approach, and report the
range and spread of the resulting by-type CAS values below.

We adopt four broad morphological types: Early, Late, Irregular, and
Other.  The classification is based largely on the $z_{850}$ data,
with cross-examination of the $B_{435}$, $V_{606}$, and $i_{775}$
imaging for the colors and multi-wavelength structure of features.
Since a primary goal of this work is to make a census of EROs that may
be present-day elliptical progenitors, we use a conservative
definition for Early types, excluding galaxies with any definite
evidence of a prominent disk or dust-band.  An isolated environment
was not a requirement, so included are galaxies that are clearly
interacting or merging.  The Late-type category consists of galaxies
with some regular, disky component, independent of the bulge fraction
-- which can be quite large.  Irregular-types are the Magellanic-type
galaxies, whereas the Other type predominantly includes low surface
brightness galaxies and chain galaxies with multiple star-formation
sites.  Two examples of each type are given in
Fig.~\ref{fig:montage}.

The classifications were done independently by four authors (SC, PE,
RL, LAM), with the consolidation done by LAM.  The overall agreement
was very good, the greatest differences arising in the use of
galaxies' central concentration as a criterion for several cases of
morphologically Early-type galaxies.  Since it became apparent that
many EROs appear to be in dynamical disequilibrium, potentially
softening the sharpness of cores, concentration as a criterion was
relaxed.

For both Early- and Late-type EROs, the mean values and typical
dispersions of the CAS parameters, are $C_{\rm E}\approx C_{\rm L}
\approx 3.1\pm0.5$, and $A_{\rm E}\approx A_{\rm L}
\approx0.25\pm0.1$.  The Irregular- and Other-type EROs are both less
concentrated (at $C_{\rm I}\approx C_{\rm O}\approx 2.5\pm0.3$) and
more asymmetric (at $A_{\rm I}\approx A_{\rm O}\approx 0.3\pm0.2$).

\begin{figure*}
\begin{center}
\resizebox{0.47\textwidth}{!}{\includegraphics{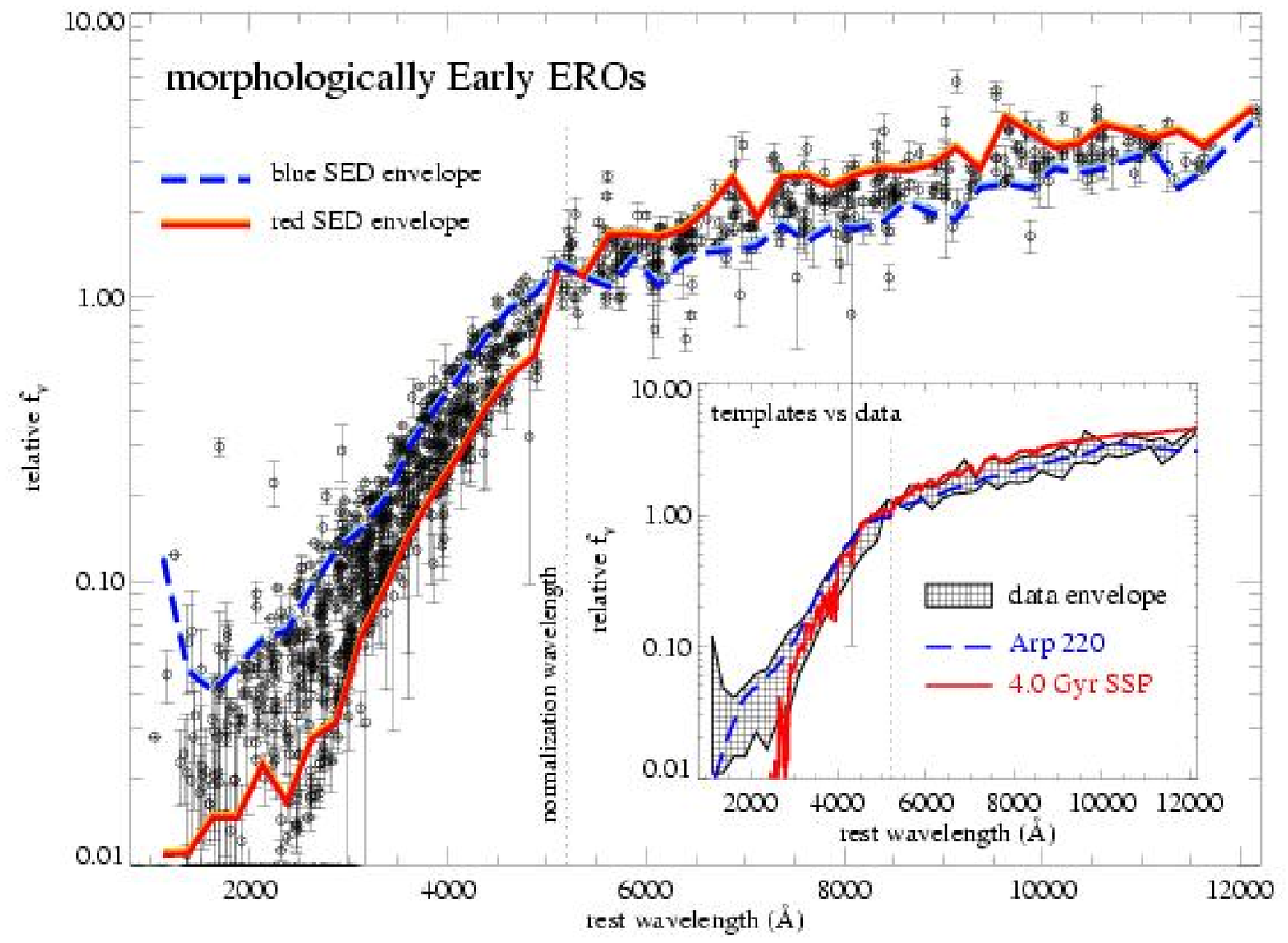}}
\resizebox{0.47\textwidth}{!}{\includegraphics{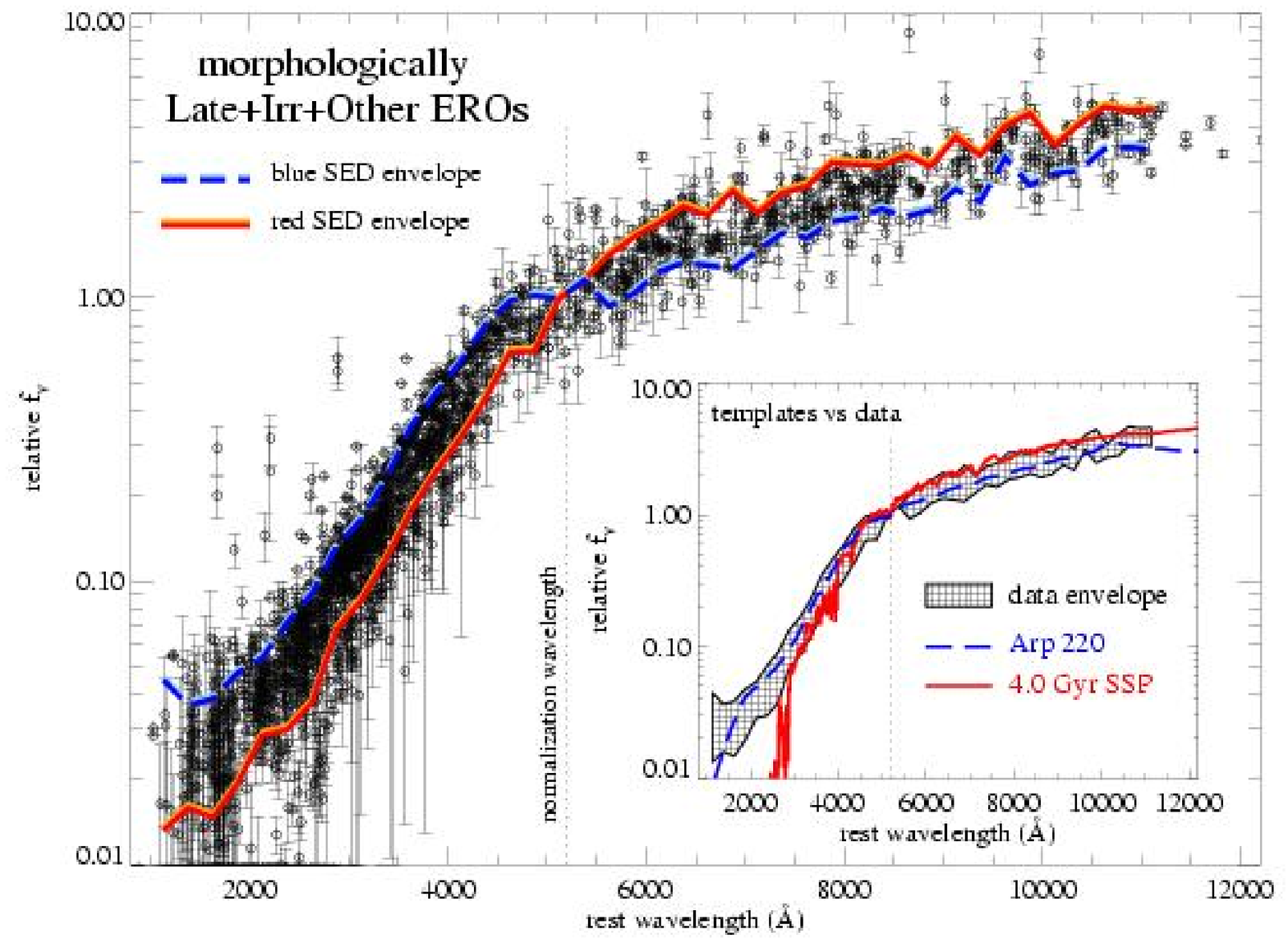}} 
\caption{\label{fig:restsed} The rest-frame SEDs for $K_s<22$,
$R-K_s>3.35$ EROs, Early at left; the combined set of Late, Irregular,
and Other types combined, at right.  These are constructed from the
broadband magnitudes, shifted to the restframe via photometric and
spectroscopic redshifts.  All SEDs are normalized to $f_{\nu}=1$ at
5200\,\AA, with the precise normalization value for each object
determined via interpolation on the best-fitting SED from
\citet{mobasher:03}. All points are shown with their photometric
uncertainty -- some of the error bars are smaller than the points.  At
each wavelength bin, a bi-weight mean and dispersion are calculated;
the blue and red ``envelope'' SEDs reflect approximately the mean
bluest and reddest SEDs in each sample.  These can then be compared
against two representative ERO templates, shown in the inset plot: the
Arp 220 SED \citep[as given in][]{elbaz:02}; and a 4.0\,Gyr old
$Z=Z_{\odot}$ simple stellar population from the v.2000 \citet{bc:93}
models.}
\end{center}
\end{figure*}

\section{Results\label{sec:results}}

The full ERO dataset comprises $\sim10$\% of all $K_s$-selected
galaxies at $K_s=20$--$22$.  The relative fractions of EROs by
morphological type are presented in Table~\ref{tab:erofrac} for
several limiting magnitudes and color limits.  We find the relative
fraction of Early-types does not depend strongly on the selection
criteria, staying at $33-44$\% of the total numbers. Late-types,
however, dominate the population at brighter magnitudes (with 55\% at
$K_s<21.0$, $R-K_s>3.35$), and comprise much smaller fractions (23\%)
at the faintest, reddest limit (see also $I_{F814W}$-band WFPC2-based
analyses by \citealt*{yan:03} and \citealt{gilbank:03}, which find
similar relative fractions).  The redshift distributions of Early- and
Late-types are very similar (Fig.~\ref{fig:nzt}).  The combined
fraction of Irregular- and Other-types increases dramatically with
both limiting magnitude and color threshold; these galaxies also tend
to have higher photometric redshifts, with a median of $z_{\rm
med}\approx1.5$.  Based on the observed redshift distributions,
comoving space densities are calculated and presented in
Table~\ref{tab:erodens}.  The volume probed by the full sample is
$\sim4.4\times10^{5}$\,$h_{70}^{-3}$\,Mpc$^3$, for a total comoving
space density of
$n=(6.40\pm0.39)\times10^{-4}$\,$h_{70}^3$\,Mpc$^{-3}$, and the cosmic
variance uncertainty is therefore $\sigma_{\rm cv}\approx15$\%
\citep{rss:03}.

To explore whether there is a useful correspondence between
morphological type and colors, we calculate the rest-frame SED of each
ERO, normalized to $f_{\nu}=1$ (arbitrary units) at 5200\,\AA.  In
Fig.~\ref{fig:restsed} we plot all data points for two groupings of
morphological types, with the corresponding mean ``blue'' and ``red''
SEDs, as described in the caption.  The strong break around
$4000$\,\AA\ follows from the red-color selection of EROs.  It is
interesting that the \emph{range} of SED shapes is extremely similar
for all EROs.  This is reflected also in the formal SED-fitting done
for calculating photometric redshifts in \citet{mobasher:03}.  Even
the reddest normal starburst SEDs \citep[drawn from][]{kinney:96} are
too blue to fit EROs, so the EROs tend strongly to be best-fit by the
earliest (elliptical) SED templates.  As seen in the inset of
Fig.~\ref{fig:restsed}, the empirical SED of the ultraluminous
infrared galaxy (ULIRG) Arp\,220 \citep[as given in][]{elbaz:02} is
sufficiently red to match ERO colors, and is coincidentally remarkably
similar to the SED of an old, passively-evolved, unreddened system.
Many of the Late-type galaxies have large bulges, and so their SEDs
may naturally be dominated by old stellar populations.  Even the
Irregular and Other morphological types alone, which may arguably be
candidates for ULIRGs, span the same range of SED shapes shown in
Fig.~\ref{fig:restsed}.  We conclude that in this wavelength-range,
color information is insufficient to \emph{generally} distinguish
between clearly distinct morphological types of EROs.

\section{Conclusions\label{sec:conc}}

Samples of color-selected EROs consist of several distinct galaxy
types for all magnitude ranges and color-cuts.  All morphological
types are found over very similar redshift ranges, $z_{\rm
med}\approx1.2$, with the more morphologically-disturbed (Irregular
and Other) types tending to somewhat higher redshifts.  This may be
due to a combination of a morphological $k$-correction selection
effects (where at redshifts above $z\approx1.3$ galaxy bulges may be
less apparent in the $z_{850}$ data); and a hint of a population of
real, ``train-wreck'' galaxies, seen at redshifts closer to
$z\approx2$.  With the present data, it is not possible to distinguish
between these possibilities from the photometry alone, as seen by the
grossly-similar rest-frame SEDs of all morphological types.  This does
suggest that those EROs that \emph{are} perhaps ``dusty starbursts,''
have SEDs that are more similar to local ULIRGs (such as Arp\,220),
rather than (mere) starburst galaxies.

The comoving space density of morphologically Early-type EROs ($n_{\rm
E}\approx2.5\times10^{-4}\,h_{70}^3\,{\rm Mpc}^{-3}$) is quite
comparable to present-day giant Ellipticals, $n_{\rm
gE,z=0}\approx2.1\times10^{-4}\,h_{70}^3\,{\rm Mpc}^{-3}$ (c.f. MS02).
This suggests that the $z\approx1.2$ Early-type galaxies identified
within the ERO population may in fact correspond to the majority of
the most massive ellipticals today, pointing to an epoch of assembly,
as well as star-formation, at redshifts well beyond $z>2$.

\acknowledgments 

We acknowledge discussion at many stages of this work with
D.~Alexander and F.~Bauer, and comments by L. Yan.  The work of DS was
carried out at JPL, Caltech, under a contract with NASA. Support for
this work, part of the \emph{Space Infrared Telescope Facility
(SIRTF)} Legacy Science Program, was provided by NASA through contract
number 1224666 issued by the Jet Propulsion Laboratory, California
Institute of Technology under NASA contract 1407.

\bibliographystyle{apj}

\begin{thebibliography}{35}
\expandafter\ifx\csname natexlab\endcsname\relax\def\natexlab#1{#1}\fi

\bibitem[{{Abraham} {et~al.}(1996){Abraham}, {Tanvir}, {Santiago},
    {Ellis}, {Glazebrook}, \& {van den Bergh}}]{abraham:96}
{Abraham}, R.\,G., {Tanvir}, N.\,R., {Santiago}, B.\,X., {Ellis},
    R.\,S.,  {Glazebrook}, K.,  \& {van den Bergh}, S. 1996, MNRAS,
    279, L47

\bibitem[{{Alexander} {et~al.}(2002){Alexander}, {Vignali}, {Bauer}, {Brandt},
  {Hornschemeier}, {Garmire}, \& {Schneider}}]{davo:02}
{Alexander}, D.\,M., {Vignali}, C., {Bauer}, F.\,E., {Brandt}, W.\,N.,
  {Hornschemeier}, A.\,E., {Garmire}, G.\,P., \& {Schneider}, D.\,P. 2002, AJ,
  123, 1149

\bibitem[{{Ben{\'{\i}}tez}(2000)}]{txitxo:00}
{Ben{\'{\i}}tez}, N. 2000, \apj, 536, 571

\bibitem[{{Bershady} {et~al.}(2000){Bershady}, {Jangren}, \&
    {Conselice}}]{bershady:00} 
{Bershady}, M.\,A., {Jangren}, A., \& {Conselice}, C.\,J. 2000, AJ, 119,
    2645 

\bibitem[{{Bolzonella} {et~al.}(2000){Bolzonella}, {Miralles}, \& {Pell{\'o}}}]{bolz:00}
{Bolzonella}, M., {Miralles}, J.-M., \& {Pell{\' o}}, R. 2000, \aap, 363, 476

\bibitem[{{Brusa} {et~al.}(2002){Brusa}, {Comastri}, {Daddi}, {Cimatti},
  {Mignoli}, \& {Pozzetti}}]{brusa:02}
{Brusa}, M., {Comastri}, A., {Daddi}, E., {Cimatti}, A., {Mignoli}, M., \&
  {Pozzetti}, L. 2002, ApJL, 581, L89

\bibitem[Bruzual \& Charlot(1993)]{bc:93} Bruzual, A.~G, \&
  Charlot, S.\ 1993, \apj, 405, 538  

\bibitem[{{Cimatti} {et~al.}(2002{\natexlab{a}}) {Cimatti}
    {et~al.}}]{cimatti:02a} {Cimatti}, A., {et~al.}
    2002{\natexlab{a}}, A\&A, 381, L68

\bibitem[{{Cimatti} {et~al.}(2002{\natexlab{b}}) {Cimatti}
    {et~al.}}]{cimatti:02b} {Cimatti}, A., {et~al.}
    2002{\natexlab{b}}, A\&A, 391, L1

\bibitem[{{Cimatti} {et~al.}(2002{\natexlab{c}}) {Cimatti}
    {et~al.}}]{cimatti:02c} {Cimatti}, A., {et~al.}
    2002{\natexlab{c}}, A\&A, 392, 395

\bibitem[Conselice(2003)]{conselice:03} Conselice, C.~J\ 2003, ApJS,
  147, 1

\bibitem[{{Daddi} {et~al.}(2000{\natexlab{a}}){Daddi}
    {et~al.}}]{daddi:00a} {Daddi}, E., {et~al.} 2000{\natexlab{a}},
    \aap, 361, 535 

\bibitem[{{Daddi} {et~al.}(2000{\natexlab{b}}){Daddi}
    {et~al.}}]{daddi:00b} {Daddi}, E., {Cimatti}, A., {Renzini},
    A. 2000{\natexlab{b}}, \aap, 362, L45 

\bibitem[{{Daddi} {et~al.}(2002){Daddi}
    {et~al.}}]{daddi:02} {Daddi}, E., {et~al.} 2002, \aap, 384, L1

\bibitem[{{Elbaz} {et~al.}(2002){Elbaz}, {Flores}, {Chanial},
  {Mirabel}, {Sanders}, {Duc}, {Cesarsky}, \& {Aussel}}]{elbaz:02}
  {Elbaz}, D., {Flores}, H., {Chanial}, P., {Mirabel}, I.\,F.,
  {Sanders}, D., {Duc}, P.-A., {Cesarsky}, C.\,J., \& {Aussel},
  H. 2002, A\&A, 381, L1

\bibitem[{{Firth} {et~al.}(2002){Firth} {et~al.}}]{firth:02} {Firth},
    A.\,E., {et~al.} 2002, MNRAS, 332, 617

\bibitem[{{Giacconi} {et~al.}(2002){Giacconi} {et~al.}}]{giac:02}
    {Giacconi}, R., {et~al.} 2002, ApJS, 139, 369

\bibitem[Giavalisco et al.(1996)]{giav:96} Giavalisco, M., Livio, M.,
Bohlin, R.\,C., Macchetto, F.\,D., \& Stecher, T.\,P. 1996, AJ, 112,
369

\bibitem[{{Giavalisco} {et~al.}(2003)}]{giav:03} {Giavalisco}, M. {et
al.} 2003, ApJL, in press

\bibitem[Gilbank et al.(2003)]{gilbank:03} Gilbank, D.~G., Smail, I.,
  Ivison, R.~J., \& Packham, C. 2003, preprint (astro-ph/0308318)

\bibitem[{{Graham} \& {Dey}(1996)}]{gd:96} {Graham}, J.\,R. \& {Dey},
A. 1996, ApJ, 471, 720

\bibitem[{{Kinney} {et~al.}(1996){Kinney}, {Calzetti}, {Bohlin},
  {McQuade}, {Storchi-Bergmann}, \& {Schmitt}}]{kinney:96} {Kinney},
  A.\,L., {Calzetti}, D., {Bohlin}, R.\,C., {McQuade}, K.,
  {Storchi-Bergmann}, T., \& {Schmitt}, H.\,R. 1996, \apj, 467, 38

\bibitem[{{Mobasher} {et al.}(2003)}]{mobasher:03} {Mobasher}, B. {et
al.} 2003, ApJL, in press

\bibitem[{{Moriondo} {et~al.}(2000){Moriondo}, {Cimatti}, \&
  {Daddi}}]{moriondo:00} {Moriondo}, G., {Cimatti}, A., \& {Daddi},
  E. 2000, A\&A, 364, 26

\bibitem[Moustakas et al.(1997)]{moustakas:97} Moustakas, L.~A.,
Davis, M., Graham, J.~R., Silk, J., Peterson, B.~A., \& Yoshii, Y.\
1997, \apj, 475, 445

\bibitem[{{Moustakas} \& {Somerville}(2002)}]{ms:02} {Moustakas},
L.\,A. \& {Somerville}, R.\,S. 2002, ApJ, 577, 1 (MS02)

\bibitem[{{Moy} {et~al.}(2003){Moy}, {Barmby}, {Rigopoulou}, {Huang},
  {Willner}, \& {Fazio}}]{moy:03} {Moy}, E., {Barmby}, P.,
  {Rigopoulou}, D., {Huang}, J.-S., {Willner}, S.\,P., \& {Fazio},
  G.\,G. 2003, A\&A, 403, 493

\bibitem[{{Pozzetti} \& {Mannucci}(2000)}]{pm:00} {Pozzetti}, L. \&
{Mannucci}, F. 2000, MNRAS, 317, L17

\bibitem[{{Roche} {et~al.}(2002){Roche}, {Almaini}, {Dunlop},
  {Ivison}, \& {Willott}}]{roche:02} {Roche}, N.\,D., {Almaini}, O.,
  {Dunlop}, J., {Ivison}, R.\,J., \& {Willott}, C.\,J. 2002, MNRAS,
  337, 1282

\bibitem[{{Smail} {et~al.}(2002){Smail}, {Owen}, {Morrison}, {Keel},
  {Ivison}, \& {Ledlow}}]{smail:02} {Smail}, I., {Owen}, F.\,N.,
  {Morrison}, G.\,E., {Keel}, W.\,C., {Ivison}, R.\,J., \& {Ledlow},
  M.\,J. 2002, ApJ, 581, 844

\bibitem[{{Somerville} {et~al.}(2003)}]{rss:03} {Somerville},
R.\,S. {et~al.} 2003, ApJL, in press

\bibitem[{{Spinrad} {et~al.}(1997){Spinrad}, {Dey}, {Stern}, {Dunlop},
  {Peacock}, {Jimenez}, \& {Windhorst}}]{spinrad:97} {Spinrad}, H.,
  {Dey}, A., {Stern}, D., {Dunlop}, J., {Peacock}, J., {Jimenez}, R.,
  \& {Windhorst}, R. 1997, ApJ, 484, 581

\bibitem[{{Wehner} {et~al.}(2002)}]{wehner:02} {Wehner}, E.\,H.,
{Barger}, A.\,J., \& {Kneib}, J.\,P. 2002, ApJ, 577, L83

\bibitem[{{Willott} {et~al.}(2001)}]{willott:01} {Willott}, C.\,J.,
{Rawlings}, S., \& {Blundell}, K.\,M. 2001, MNRAS, 324, 1

\bibitem[{{Yan} \& {Thompson}(2003)}]{yan:03} {Yan}, L. \& {Thompson},
D. 2003, ApJ, 586, 765

\end{thebibliography}

\clearpage


\begin{center}
\begin{table}
\begin{center}
\scriptsize
\caption{\label{tab:erofrac}\scriptsize ERO fractions (\%) by
morphological type$^{\ast}$}
\begin{tabular}{c c ccc c ccc}
 \multicolumn{1}{c}{} & 
 \multicolumn{1}{c}{} & 
 \multicolumn{3}{c}{$K_s<21.0$} &
 \multicolumn{1}{c}{} & 
 \multicolumn{3}{c}{$K_s<22.0$} \\
\cline{3-5}\cline{7-9}
 & $R-K_{s}>$ & 
 3.35 & 3.65 & 4.35 &
 & 
 3.35 & 3.65 & 4.35 \\
\hline\hline
\emph{early}         & &  36 &  41  &  44 & & 37 & 38 & 33 \\   
\emph{late}          & &  55 &  52  &  39 & & 38 & 32 & 23 \\   
\emph{irregular}     & &  5.1 & 5.5 &  17 & & 11 & 12 & 23 \\   
\emph{other}         & &  2.5 & 1.4 & 0.0 & & 15 & 17 & 20 \\   
\cline{3-5}\cline{7-9}
{\bf total number}$^\dagger$   & & {\bf 118} & {\bf 73}& {\bf 18}& & {\bf 275}& {\bf 185} & {\bf 60}\\   
\hline
\end{tabular}
\end{center}
\smallskip
\scriptsize
$^{\ast}$: The relative fractions are given in percent (\%).\\
$^{\dagger}$: Total number of EROs with shown criteria.
\end{table}
\end{center}

\begin{center}
\begin{table}
\begin{center}
\scriptsize
\caption{\label{tab:erodens}\scriptsize ERO space densities [$\times
  10^{-4}$\,$h_{70}^3$\,Mpc$^{-3}$] by morphological type$^{\ast}$}
\begin{tabular}{c ccc c ccc}
 \multicolumn{1}{c}{} & 
 \multicolumn{3}{c}{$K_s<21.0$} &
 \multicolumn{1}{c}{} & 
 \multicolumn{3}{c}{$K_s<22.0$} \\
\cline{2-4}\cline{6-8}
 & $R-K_{s}>3.35$  & $R-K_{s}>3.65$  & $R-K_{s}>4.35$ &
 & $R-K_{s}>3.35$  & $R-K_{s}>3.65$  & $R-K_{s}>4.35$ \\
\hline\hline
\emph{early}          & $1.10 \pm 0.17$ & $0.67 \pm 0.12$ & $0.09 \pm 0.03$ & & $2.49 \pm 0.25$ & $1.59 \pm 0.19$ & $0.29 \pm 0.07$ \\
\emph{late}           & $1.45 \pm 0.18$ & $0.86 \pm 0.14$ & $0.14 \pm 0.05$ & & $2.36 \pm 0.23$ & $1.37 \pm 0.18$ & $0.25 \pm 0.07$ \\
\emph{l+i+o}$^{\dagger}$ & $1.67 \pm 0.19$ & $0.98 \pm 0.15$ & $0.17 \pm 0.05$ & & $3.84 \pm 0.29$ & $2.53 \pm 0.24$ & $0.72 \pm 0.11$ \\
\emph{all}            & $2.94 \pm 0.27$ & $1.64 \pm 0.19$ & $0.26 \pm 0.06$ & & $6.40 \pm 0.39$ & $4.11 \pm 0.30$ & $1.03 \pm 0.13$ \\
\hline
\end{tabular}
\end{center}
\smallskip
\scriptsize
$^{\ast}$: The space density of EROs with specified $R-K_s$ and $K_s$
  thresholds, in units of $10^{-4}$\,$h_{70}^3$\,Mpc$^{-3}$, with
  Poisson uncertainties. \\
$^{\dagger}$: Results for the combined numbers of Late-, Irregular-,
  and Other-type EROs.\\
\end{table}
\end{center}

\end{document}